\documentclass[letterpaper,10pt]{article}

\usepackage{graphicx}
\usepackage{amsmath}
\usepackage{amssymb}
\usepackage{psfrag}
\newcommand{\dirac}{{\slash \negthinspace \negthinspace \negthinspace \nabla}}
\newcommand{\dd}{\textrm{d}}
\newcommand{\im}{{\mathbb{I}}{\mathrm{m}}}

\title{Quasinormal modes and stability of a five-dimensional dilatonic black hole}

\author{A.~L\'opez-Ortega\thanks{alopezo ``at'' ipn.mx} \\
		Centro de Investigaci\'on en Ciencia Aplicada y Tecnolog\'{\i}a Avanzada. \\
	      Unidad Legaria. Instituto Polit\'ecnico Nacional. \\
             Calzada Legaria \# 694. Colonia Irrigaci\'on. Delegaci\'on Miguel Hidalgo. \\
	      M\'exico, D.\ F., M\'exico. \\
	      C.\ P.\  11500\\ 
               }

\begin{document}

\maketitle

\begin{abstract}

We exactly calculate the quasinormal frequencies of the electromagnetic and Klein-Gordon perturbations propagating in a five-dimensional dilatonic black hole. Furthermore we exactly find the quasinormal frequencies of the massive Dirac field. Using these results we study the linear stability of this black hole. We compare our results for the quasinormal frequencies and for the linear stability of the five-dimensional black hole with those already published.

\end{abstract}



RUNNING HEAD: Quasinormal modes and stability of a black hole

\section{Introduction}
\label{sect 1}

In General Relativity and related areas, the quasinormal modes (QNMs) of a black hole are physical quantities whose computation is useful because they depend on the physical parameters that characterize the spacetime and the classical field. Thus if we know the properties of the field, by measuring its quasinormal frequencies (QNF) we infer the values corresponding to several physical quantities of the black hole, (see Refs.\ \cite{Kokkotas:1999bd} for comprehensive reviews on QNMs).

The QNMs of the gravitational perturbations propagating in several four-dimensional black holes were calculated because we expect these modes will be significant in gravitational wave astronomy \cite{Kokkotas:1999bd,Ferrari:2007dd}. Also the QNMs are useful to study the linear stability of the black holes, because if we find modes satisfying the boundary conditions of the QNMs and with amplitudes increasing in time, then the black hole is linearly unstable \cite{Konoplya:2007jv,Becar:2007hu}. Also, recently the QNMs have been helpful in other research lines such as the AdS/CFT correspondence of String Theory \cite{Horowitz:1999jd,Birmingham:2001pj}, and Hod's conjecture in Quantum Gravity \cite{Hod:1998vk}.

We know several three-dimensional and two-dimensional spacetimes whose QNF were calculated exactly, for example, the three-dimensional BTZ black hole \cite{Birmingham:2001pj,Cardoso:2001hn,Crisostomo:2004hj}, the three-dimensional black hole of the Einstein-Maxwell-Dilaton with cosmological constant theory \cite{Fernando:2003ai,Fernando:2008hb,LopezOrtega:2005ep}, the three-dimensional de Sitter spacetime \cite{Lopez-Ortega:2006ig,Du:2004jt}, and the two-di\-men\-sion\-al uncharged dilatonic black hole \cite{Becar:2007hu}.

Generally for four-dimensional and  $D$-dimensional black holes ($D>4$) their QNF are calculated by using approximate analytical methods or numerical techniques. See reviews \cite{Kokkotas:1999bd}, \cite{Ferrari:2007dd}, and Refs.\ \cite{Konoplya:2007jv}, \cite{Horowitz:1999jd}, \cite{Vishveshwara:1970} (this list is not exhaustive). Nevertheless there are some higher dimensional backgrounds whose QNF were computed exactly. We know the $D$-dimensional massless topological black hole \cite{Aros:2002te,Birmingham:2006zx,LopezOrtega:2007vu}, the $D$-dimensional Nariai spacetime \cite{LopezOrtega:2007vu,Vanzo:2004fy}, the $D$-dimensional de Sitter spacetime \cite{Du:2004jt,Lopez-Ortega:2006my,LopezOrtega:2007sr,Natario:2004jd}, the BTZ black string \cite{Liu:2008ds}, and the five-dimensional uncharged dilatonic black hole \cite{Becar:2007hu}.

Recently the linear stability of higher dimensional spacetimes has attracted a lot of attention, see for example Refs.\ \cite{Konoplya:2007jv}, \cite{Becar:2007hu}, \cite{Ishibashi:2003ap}-\cite{Moura:2006pz} (this list is not exhaustive). Becar et al.\ in Ref.\ \cite{Becar:2007hu} exactly calculated the QNF of a non-minimally coupled to gravity Klein-Gordon field propagating in a five-dimensional dilatonic black hole to study the linear stability. This five-dimensional dilatonic black hole is one of a class of spacetimes for which their statistical entropies were calculated in Refs.\ \cite{Strominger:1996sh}-\cite{Hyun:1997jv}. We note that in these references some results of String Theory were used in order to compute the statistical entropy of the black holes. Furthermore, exploiting duality relations in Ref.\ \cite{Teo:1998kp} was computed the statistical entropy of the two-dimensional black hole given in Ref.\ \cite{McGuigan:1991qp} from the result for the five-dimensional dilatonic black hole that we study in the present paper. 

We think this five-dimensional dilatonic black hole as a simple higher dimensional generalization of the uncharged two-dimensional black hole of Ref.\ \cite{McGuigan:1991qp}, since its metric is the direct product of the metrics for the previously mentioned two-dimensional black hole and the three-dimensional sphere. 

From their result on the QNF of the Klein-Gordon field propagating in the five-dimensional black hole, Becar et al.\ in Ref.\ \cite{Becar:2007hu} studied its linear stability and inferred that this black hole is linearly stable under Klein-Gordon fields. Thus in Ref.\ \cite{Becar:2007hu} was deduced that the extra spatial dimensions stabilize the five-dimensional dilatonic black hole, because in Refs.\ \cite{Becar:2007hu}, \cite{AzregAinou:1999kb}, \cite{Frolov:2000jh}, \cite{Zelnikov:2008rg} was shown that the two-dimensional black hole of Ref.\ \cite{McGuigan:1991qp} is linearly unstable under Klein-Gordon and gravitational perturbations. 

In this paper we exactly calculate the QNF for the scalar and vector type electromagnetic fields and for the massive Klein-Gordon and Dirac fields propagating in the five-dimensional dilatonic black hole already mentioned. Using these results we study the linear stability of this black hole. Our conclusion about the linear stability of the five-dimensional black hole is different from that by Becar, et al.\ because for these fields we find QNMs with amplitudes increasing in time. We discuss in depth these results in the main body of the paper. Here we only comment that our results are an extension and revision of those by Becar, et al.\ published in Ref.\ \cite{Becar:2007hu}.

We organize this paper as follows. In Sect.\ \ref{sect 2} we exactly calculate the QNF of the scalar and vector type electromagnetic fields and of the minimally coupled Klein-Gordon field propagating in the five-dimensional black hole that was studied in Ref.\ \cite{Becar:2007hu}. Exploiting these results we discuss its stability. Also we review the results of Ref.\ \cite{Becar:2007hu} on the QNF of the non-minimally coupled to gravity Klein-Gordon field and on the linear stability of the five-dimensional black hole. In Sect.\ \ref{sect 3} we extend our previous analysis to include the massive Dirac field. In Sect.\ \ref{sect 4} we discuss our main results. Finally in Appendix \ref{app 1} we make some comments on the results of Ref.\ \cite{Becar:2007hu} about the QNF of the Klein-Gordon field propagating in the uncharged two-dimensional dilatonic black hole of Ref.\ \cite{McGuigan:1991qp}.

\section{QNMs of boson fields}
\label{sect 2}

In this paper we study the spherically symmetric five-dimensional dilatonic black hole whose line element we write as \cite{Becar:2007hu,Teo:1998kp} 
\begin{equation} \label{eq: metric dilatonic} 
{\rm d} s^2 = (1 - \textrm{e}^{-x})\, {\rm d} t^2 - \frac{ {\rm d} x^2}{1 -  \textrm{e}^{-x} } -  r_0^2 \,{\rm d} \Sigma_{3}^2 ,
\end{equation} 
where $x \in (0, + \infty)$, $r_0$ is a constant and $\dd \Sigma^2_3$ is the line element of a unit three-dimensional sphere. Notice that the metric (\ref{eq: metric dilatonic}) is the direct product of the metrics for the asymptotically flat two-dimensional uncharged black hole of Ref.\ \cite{McGuigan:1991qp} and the three-dimensional sphere of radius $r_0$, also note that these two metrics are decoupled. Thus we think that the five-dimensional black hole (\ref{eq: metric dilatonic}) is a simple higher dimensional generalization of the asymptotically flat two-dimensional black hole of Ref.\ \cite{McGuigan:1991qp}. For more details about this black hole see Refs.\ \cite{Becar:2007hu}, \cite{Teo:1998kp}.

Notice that there are no many papers in which the QNF of dilatonic black holes are computed, we know Refs.\ \cite{Becar:2007hu}, \cite{Fernando:2003ai}, \cite{Fernando:2008hb}, \cite{LopezOrtega:2005ep}, \cite{Ferrari:2000ep}. Exact results are reported in Refs.\ \cite{Becar:2007hu}, \cite{Fernando:2003ai}, \cite{Fernando:2008hb}, \cite{LopezOrtega:2005ep}. In the present paper we extend some of the results given in these references. 

The QNMs of the five-dimensional dilatonic black hole (\ref{eq: metric dilatonic}) are defined as the solutions to the equations of motion that are purely ingoing at the event horizon and purely outgoing at infinity \cite{Becar:2007hu}. First in this section we calculate the QNF of the scalar and vector type electromagnetic fields and of the minimally coupled massless Klein-Gordon field. Note that the results of this paper are an extension of those already published by Becar, et al.\ in Ref.\ \cite{Becar:2007hu} for the non-minimally coupled to gravity Klein-Gordon field. Also in this section we review the published results of Ref.\ \cite{Becar:2007hu} on the QNF of the Klein-Gordon field and on the linear stability of the five-dimensional black hole (\ref{eq: metric dilatonic}).

We notice that the five-dimensional black hole (\ref{eq: metric dilatonic}) is uncharged. As is well known, in a $D$-dimensional spherically symmetric spacetime it is possible to simplify the equations of motion for test electromagnetic fields to partial differential equations in two variables \cite{Kodama:2003kk,Kodama:2003jz,Ishibashi:2003jd}. We also obtain the same type of differential equation for the minimally coupled massless Klein-Gordon field. In the five-dimensional dilatonic black hole (\ref{eq: metric dilatonic}), Eqs.\ (53) and (67) of Ref.\ \cite{Ishibashi:2003jd} reduce to the partial differential equations 
\begin{equation} \label{eq: partial differential fields}
 \left[ \square^{2}_{2D} - \alpha_E^2 \right] \phi(x,t) = 0,
\end{equation} 
where the quantities $\alpha_E^2$ are equal to\footnote{Owing to the coupling with the perturbations of the dilaton field we do not calculate the QNF for the three types of gravitational perturbations because we cannot simplify their equations of motion to  partial differential equations similar to Eq.\ (\ref{eq: partial differential fields}).}
\begin{align}
\alpha_E^2 = & \left\{ \begin{array}{l} \tfrac{l(l+2)}{r_0^2}, \, \, \, \, \, \, \textrm{for massless Klein-Gordon field and scalar} \\ 
\quad \qquad \,\, \, \, \,  \textrm{electromagnetic perturbation}, \\
\tfrac{(l+1)^2}{r_0^2} , \, \, \, \, \, \, \textrm{for vector electromagnetic perturbation, }  \end{array} \right.
\end{align}   
the symbol $\square_{2D}^2$ denotes the d'Alembertian in the two-dimensional $(t,x)$ sector of the line element (\ref{eq: metric dilatonic}) and the quantity $\phi$ is related to the gauge invariant quantities appropriate to the type of electromagnetic perturbation (see Refs.\ \cite{Kodama:2003kk}-\cite{Ishibashi:2003jd} for more details).

Taking the quantity $\phi$ as
\begin{equation} \label{eq: phi ansatz}
 \phi (x,t)= \textrm{e}^{-i \omega t} R(x),
\end{equation}  
we get that Eq.\ (\ref{eq: partial differential fields}) reduces to the ordinary differential equation
\begin{equation} \label{eq: radial EM GRAV 1}
  \frac{\dd }{\dd x}\left((1 - \textrm{e}^{-x})\frac{\dd R}{\dd x}\right) + \frac{\omega^2 R}{1 - \textrm{e}^{-x}} - \alpha_E^2 R  = 0.
\end{equation}  
To solve this equation we make the change of variable (as in Ref.\ \cite{Becar:2007hu})
\begin{equation} \label{eq: z coordinate dilatonic}
 z = 1 - \textrm{e}^{-x},
\end{equation} 
thus $z \in (0,1)$, to find that Eq.\ (\ref{eq: radial EM GRAV 1}) becomes
\begin{align} \label{eq: dilatonic EM GRAV}
 \frac{\dd^2 R}{\dd z^2}  + \left(\frac{1}{z}- \frac{1}{1-z} \right) \frac{\dd R}{\dd z} + \frac{\omega^2 R}{z^2 (1-z)^2} - \frac{\alpha_D^2 R}{z(1-z)^2}  = 0.
\end{align}

The last equation is similar to Eq.\ (8) of Ref.\ \cite{Becar:2007hu}. Thus using the same method of Ref.\ \cite{Becar:2007hu} we shall solve Eq.\ (\ref{eq: dilatonic EM GRAV}). First we make the ansatz 
\begin{equation} \label{eq: Em GRAV ansatz}
 R(z) = z^C (1-z)^B Q(z),
\end{equation} 
where
\begin{align} \label{eq. B C EM GRAV }
C = & \left\{ \begin{array}{l} i \omega   \\ \\ - i \omega \end{array}\right. ,\qquad \quad
B =  \left\{ \begin{array}{l} i \sqrt{\omega^2 - \alpha_E^2}  \\ \\ - i \sqrt{\omega^2 - \alpha_E^2}   \end{array}\right. , 
\end{align}
to get that the function $Q$ is a solution of the hypergeometric differential equation \cite{b:DE-books}
\begin{equation} \label{e: hypergeometric differential equation}
 z(1-z) \frac{\dd^2 Q}{\dd z^2} + (c - (a +b + 1)z)\frac{\dd Q}{{\rm d}z} - a b\, Q   = 0,
\end{equation}   
where the quantities $a$, $b$, and $c$ are equal to
\begin{align} \label{eq: constants hypergeometric EM GRAV}
a & = B + C + 1, \nonumber \\
b & = B + C ,  \nonumber \\  
c & = 2 C + 1 .   
\end{align}
Therefore if the quantity $c$ is not an integer then the radial function $R$ is given by
\begin{align} \label{eq: 12}
 R = &(1-z)^{B}  \left[ \mathbb{D} \, z^{ i \omega } {}_{2}F_{1}(a,b;c;z) \right.  \nonumber \\
&\left.  + \mathbb{E} \, z^{ - i \omega} {}_{2}F_{1}(a-c+1,b-c+1;2-c;z) \right],
\end{align} 
where $\mathbb{D}$ and $\mathbb{E}$ are constants. 

It is convenient to note the following facts about the different coordinates that we use in the black hole (\ref{eq: metric dilatonic}). The tortoise coordinate for this spacetime is \cite{Becar:2007hu}
\begin{equation}
 r_* = \int \frac{\dd x}{1 - \textrm{e}^{-x}} = \ln(\textrm{e}^x - 1),
\end{equation} 
where $r_* \in (-\infty, + \infty)$, $r_* \to - \infty$ near the event horizon and $r_* \to + \infty $ at infinity. Also from the definition given in formula (\ref{eq: z coordinate dilatonic}) for the coordinate $z$ we get
\begin{equation}
 z = \frac{\textrm{e}^{r_*}}{\textrm{e}^{r_*} + 1},
\end{equation} 
and therefore 
\begin{align} \label{eq: dilatonic relations coordinates}
 \textrm{as } \quad & r_* \to - \infty,   & z \approx & \,\textrm{e}^{r_*},  \nonumber \\
\textrm{as } \quad & r_* \to + \infty,  &  1-z \approx &\,\textrm{e}^{-r_*} .  
\end{align}

To calculate the QNF of the fields studied in this section we choose $C = i \omega $ and $B = i \sqrt{\omega^2 - \alpha_E^2}$.  From formulas (\ref{eq: dilatonic relations coordinates}), for the function $R$ of Eq.\ (\ref{eq: 12}) we find that near the horizon the first term is a purely outgoing wave, whereas the second term is a purely ingoing wave. To satisfy the boundary condition of the QNMs near the horizon we must take  $\mathbb{D}=0$. Thus 
\begin{align} \label{eq: radial EM GRAV}
 R & = \mathbb{E} \, z^{- i\omega} (1-z)^{B} \, {}_{2}F_{1}(a-c+1,b-c+1;2-c;z) \nonumber \\
& = \mathbb{E} \, z^{ - i\omega} (1-z)^{B} \, {}_{2}F_{1}(\alpha,\beta;\gamma;z).
\end{align}

At this point we recall that if the quantity $c-a-b$ is not an integer then the hypergeometric function ${}_{2}F_{1}(a,b;c;u)$ satisfies \cite{b:DE-books}
\begin{align} \label{e: hypergeometric property z 1-z}
{}_2F_1(a,b;c;u) &= \frac{\Gamma(c) \Gamma(c-a-b)}{\Gamma(c-a) \Gamma(c - b)} {}_2 F_1 (a,b;a+b+1-c;1-u) \nonumber \\
&+ \frac{\Gamma(c) \Gamma( a + b - c)}{\Gamma(a) \Gamma(b)} (1-u)^{c-a -b} {}_2F_1(c-a, c-b; c + 1 -a-b; 1 -u).
\end{align}  
Using this formula and Eqs.\ (\ref{eq: dilatonic relations coordinates}) we find that near infinity the radial function $R$ simplifies to
\begin{align} \label{eq: EM GRAV near infinity}
 R \approx & \frac{\Gamma(\gamma) \Gamma(\gamma - \alpha - \beta)}{\Gamma(\gamma - \alpha) \Gamma(\gamma - \beta)}  \textrm{e}^{- i \sqrt{\omega^2 - \alpha_E^2} r_* }   + \frac{\Gamma(\gamma) \Gamma( \alpha + \beta - \gamma)}{\Gamma(\alpha) \Gamma(\beta)} \textrm{e}^{i \sqrt{\omega^2 - \alpha_E^2} r_*} .
\end{align}

According to Refs.\ \cite{Ohashi:2004wr} the first term is an ingoing wave and the second term an outgoing wave. Thus to fulfill the boundary conditions of the QNMs we must impose the condition 
\begin{equation}
 \gamma - \alpha = - n , \qquad \textrm{or} \qquad \gamma - \beta = -n .
\end{equation}  
From these equations we get that for the electromagnetic and massless Klein-Gordon perturbations their QNF are equal to
\begin{align} \label{eq: QNF GRAV EM}
 \omega = - \frac{i}{2n}\left( n^2 - \alpha_E^2 \right), \qquad 
 \omega = - \frac{i}{2(n+1)}\left( (n+1)^2 - \alpha_E^2 \right) ,
\end{align} 
where $n=1,2,3,\dots$, for the first set of QNF and $n=0,1,2,3,\dots$, for the second set of QNF.

As the time dependence is of the form $\textrm{exp}(-i \omega t)$ (see formula (\ref{eq: phi ansatz})), in order to have stable QNMs we need that $\im(\omega) < 0$. Thus for $\alpha_E^2 > n^2$ or $\alpha_E^2 > (n+1)^2$ the amplitudes of the QNMs increase with time. This fact shows that in the five-dimensional black hole (\ref{eq: metric dilatonic}) there are unstable QNMs, (for the fields studied in this section at least the fundamental mode is unstable). This result about the linear stability of the black hole (\ref{eq: metric dilatonic}) is different from that obtained by Becar, et al.\ in Ref.\ \cite{Becar:2007hu}. They inferred that the five-dimensional black hole (\ref{eq: metric dilatonic}) is linearly stable by studying only the Klein-Gordon field with non-minimal coupling to gravity (see below).

We note that the QNF (\ref{eq: QNF GRAV EM}) are purely imaginary as those already found in Refs.\ \cite{Fernando:2003ai}, \cite{Fernando:2008hb}, \cite{LopezOrtega:2005ep}, \cite{LopezOrtega:2007sr}, \cite{Saavedra:2005ug} for several fields and spacetimes (see also the QNF (\ref{eq: QNF Klein Gordon}) below and expressions (\ref{eq: QN frequencies dilatonic}) and (\ref{eq: QN frequencies dilatonic R2}) of Sect.\ \ref{sect 3}). Moreover notice that for $\alpha_E \neq 0$ the QNF (\ref{eq: QNF GRAV EM}) depend on the mode number $n$ in the form $1/n$ and $1/(n+1)$ ($1/n$ for large $n$).

Furthermore the QNF given in formula (\ref{eq: QNF GRAV EM}) have the same mathematical form that the limit $n \to + \infty$ of the QNF for the four-dimensional Schwarzschild black hole given by Eq.\ (12) of Ref.\ \cite{Liu-Mashhoon-1996}. We cannot explain this coincidence in the mathematical form of the QNF (\ref{eq: QNF GRAV EM}) and those of Ref.\ \cite{Liu-Mashhoon-1996}. We only note that in Ref.\ \cite{Liu-Mashhoon-1996} Liu and Mashhoon obtained the solutions of the radial equation in terms of the confluent hypergeometric function but in this section the solutions involve the hypergeometric function.\footnote{We remark that Eq.\ (12) of Ref.\ \cite{Liu-Mashhoon-1996} is valid in the limit $n \to + \infty$. Therefore the condition $j+\tfrac{1}{2} > (n + \tfrac{1}{2})/\sqrt{2}$ necessary for the existence of unstable QNMs cannot be satisfied. This fact agrees with the linear stability of the four-dimensional Schwarzschild background \cite{Kay:1987ax}, (in this formula $j$ is the angular mode number).}

We note that the quantity $\alpha_E$ can be equal to zero when $l=0$ (that is for the Klein-Gordon field). Thus we expect that the frequencies given in expression (\ref{eq: QNF GRAV EM}) with $\alpha_E = 0$ are QNF. Nevertheless we need to verify if the modes with these frequencies fulfill the boundary conditions of the  QNMs. Here we follow a slightly different path. We solve Eq.\ (\ref{eq: dilatonic EM GRAV}) with $\alpha_E = 0$ to disprove the existence of solutions that represent QNMs.

If $\alpha_E = 0$ then Eq.\ (\ref{eq: dilatonic EM GRAV}) reduces to
\begin{align} \label{eq: EM GRAV alpha}
 \frac{\dd^2 R}{\dd z^2}  + \left(\frac{1}{z}- \frac{1}{1-z} \right) \frac{\dd R}{\dd z} + \frac{\omega^2 R}{z^2 (1-z)^2} = 0.
\end{align}
We transform the previous equation to the form 
\begin{equation} \label{eq: Schrodinger type alpha}
 \frac{\dd ^2 R}{\dd u^2} + \omega^2 R = 0 ,
\end{equation} 
where we define the variable $u$ by 
\begin{equation}
 \frac{\dd u}{\dd z} = \frac{1}{z(1-z)},
\end{equation}  
thus
\begin{equation}
 u = \ln(z) - \ln(1-z).
\end{equation} 
Therefore the solutions of Eq.\ (\ref{eq: EM GRAV alpha}) are of the form
\begin{equation} \label{eq: R alpha zero}
 R = \mathbb{D} \,\, z^{i \omega } (1-z)^{-i \omega} + \mathbb{E} \,\, z^{-i \omega} (1-z)^{i \omega},
\end{equation} 
where $\mathbb{D}$ and $\mathbb{E}$ are constants, as before. 

From formulas (\ref{eq: dilatonic relations coordinates}) we find that the first term is an outgoing wave near the event horizon and at infinity. The second term is an ingoing wave near the horizon and at infinity. Thus the function $R$ given in formula (\ref{eq: R alpha zero}) does not satisfy the boundary conditions of the QNMs.

For $l=0$ the effective potential in the Schr\"odinger type equation (\ref{eq: Schrodinger type alpha}) is equal to zero, thus this mode of the field propagates freely in the five-dimensional black hole (\ref{eq: metric dilatonic}). Therefore in this black hole for $\alpha_E = 0$ ($l=0$) there are no QNMs.

Next, we review the already published results of Ref.\ \cite{Becar:2007hu} on the QNF of the Klein-Gordon field with non-minimal coupling to gravity, because we believe that some of its results are incorrect. The equation of motion for the non-minimally coupled to gravity Klein-Gordon field $\Phi$ is 
\begin{equation}  \label{eq: Klein-Gordon massive}
 (\square^2 - \mu^2 - \xi \mathcal{R}) \Phi = 0,
\end{equation} 
where  $\square^2$ is the d'Alembertian, $\xi$ is the coupling constant between the massive Klein-Gordon field and the scalar curvature of the spacetime (\ref{eq: metric dilatonic}) given by $\mathcal{R} = \textrm{e}^{-x} + 6/r_0^2$, and $\mu$ is the mass of the field.

To solve this equation in the five-dimensional black hole (\ref{eq: metric dilatonic}) we make the ansatz
\begin{equation} \label{eq: KG ansatz}
 \Phi(x,t,\theta_i) = R(x) \textrm{e}^{-i \omega t} Y(\theta_i),
\end{equation} 
where $Y(\theta_i)$, $i=1,2,3$, are the spherical harmonics on the three-dimensional sphere. Substituting expression (\ref{eq: KG ansatz}) into Eq.\ (\ref{eq: Klein-Gordon massive}) we find that the function $R$ is a solution of the ordinary differential equation
\begin{equation} \label{eq: radial Klein Gordon}
 (1 - \textrm{e}^{-x}) \frac{\dd^2 R }{\dd x^2} +  \textrm{e}^{-x} \frac{\dd R}{\dd x} + \frac{\omega^2 R}{1 - \textrm{e}^{-x}} - \left( \frac{l(l+2) + 6 \xi}{r_0^2} + \mu^2 \right) R - \xi \textrm{e}^{-x} R = 0.
\end{equation} 

This differential equation is similar to Eq.\ (\ref{eq: radial EM GRAV 1}) and we use the previous method to solve it. Thus making the ansatz (\ref{eq: Em GRAV ansatz}) we find that for the Klein-Gordon field with non-minimal coupling to gravity the quantities $C$ and $B$ are equal to
\begin{align} \label{eq. B C Klein Gordon}
C = & \left\{ \begin{array}{l} i \omega   \\ \\ - i \omega \end{array}\right. ,\quad \quad
B =  \left\{ \begin{array}{l} i \sqrt{\omega^2 - \frac{l(l+2) + 6  \xi}{r_0^2} - \mu^2 }  \\ \\ - i \sqrt{\omega^2 - \frac{l(l+2) + 6  \xi}{r_0^2} - \mu^2 }   \end{array}\right. , 
\end{align} 
and the function $Q$ is a solution to the hypergeometric differential equation (\ref{e: hypergeometric differential equation}) with the parameters $a$, $b$, and $c$ given by\footnote{We choose $C = - i \omega$ and $B = - i \sqrt{ \omega^2 - \frac{l(l+2) + 6  \xi}{r_0^2} - \mu^2 }$ in the following.}
\begin{align} \label{eq: constants hypergeometric Klein Gordon}
a & = B + C + \frac{1}{2} + \frac{\sqrt{1 - 4 \xi}}{2}, \nonumber \\
b & = B + C + \frac{1}{2} - \frac{\sqrt{1 - 4 \xi}}{2},  \nonumber \\
c & = 2 C + 1 .   
\end{align}

Next we impose the boundary conditions of the QNMs to obtain that for the non-minimally coupled Klein-Gordon field evolving in the five-dimensional black hole (\ref{eq: metric dilatonic}) the QNF are equal to
\begin{equation} \label{eq: QNF Klein Gordon}
 \omega = - \frac{i}{2} \left[ \frac{1}{2} \pm  \frac{\sqrt{1 - 4 \xi}}{2} + n - \frac{\frac{l(l+2) + 6  \xi}{r_0^2} + \mu^2}{\frac{1}{2} \pm  \frac{\sqrt{1 - 4 \xi}}{2} + n}    \right].
\end{equation} 
This result is different from that given in Eq.\ (44) of Ref.\ \cite{Becar:2007hu} (here we write that result in our notation)
\begin{equation}
 \omega = -\frac{i}{4} \left( 1 - \sqrt{1 - 4 \xi} - \frac{(1 + \sqrt{1 - 4 \xi} ) \mu^2 - l(l+2)  }{n^2 +  n +\xi}  + n \left( 2 - \frac{2 \mu^2 - 2l(l+2)}{n^2 +  n +\xi} \right) \right).
\end{equation}  
Also notice that in Ref.\ \cite{Becar:2007hu} only a set of QNF is reported, but we found two sets of QNF (see our expression (\ref{eq: QNF Klein Gordon}) above). 

Using the result (\ref{eq: QNF Klein Gordon}) for the QNF we discuss the linear stability of the five-dimensional black hole (\ref{eq: metric dilatonic}) under Klein-Gordon fields with non-minimal coupling to gravity. If $\xi < \tfrac{1}{4}$ the QNF (\ref{eq: QNF Klein Gordon}) are purely imaginary. In order that the amplitudes of the QNMs decay in time we need that $\im (\omega) < 0$. From expression (\ref{eq: QNF Klein Gordon}) for the QNF of the Klein-Gordon field we find that if the quantities $n$, $l$, $\xi$, and $\mu$ satisfy  
\begin{equation} 
 \left( \frac{1}{2} \pm  \frac{\sqrt{1 - 4 \xi}}{2} + n \right)^2 < \mu^2 +  \frac{l(l+2) + 6  \xi}{r_0^2},
\end{equation} 
then the amplitudes of the QNMs grow in time. Furthermore, for $\xi > \tfrac{1}{4}$ we find similar results to those obtained for $\xi < \tfrac{1}{4}$.

Our result about the linear stability of the five-dimensional black hole (\ref{eq: metric dilatonic}) is different from that of Becar, et al.\ given in Ref.\ \cite{Becar:2007hu}. They inferred that the extra three spatial dimensions stabilize the asymptotically flat two-dimensional black hole in the $(t,x)$ sector of the spacetime (\ref{eq: metric dilatonic}). This two-dimensional black hole is linearly unstable under Klein-Gordon and gravitational perturbations as was shown in Refs.\ \cite{Becar:2007hu}, \cite{AzregAinou:1999kb}, \cite{Frolov:2000jh}, \cite{Zelnikov:2008rg}. The values for the QNF given in Eqs.\ (\ref{eq: QNF GRAV EM}) and (\ref{eq: QNF Klein Gordon}) prove that for the five-dimensional black hole (\ref{eq: metric dilatonic}) there are QNMs with amplitudes increasing in time, hence there are unstable QNMs. This result points out the existence of a linear instability in the five-dimensional black hole (\ref{eq: metric dilatonic}).

To finish this section we note that Eq.\ (\ref{eq: radial EM GRAV 1}), (valid for the scalar and vector type electromagnetic fields and the minimally coupled massless Klein-Gordon field), reduces to a Schr\"odinger type equation with effective potential 
\begin{equation} \label{eq: potential EM GRAV}
 V(r_*) = \frac{\alpha_E^2 \textrm{e}^{r_*}}{\textrm{e}^{r_*} + 1} ,
\end{equation} 
thus $V=0$ when $\alpha_E = 0$, as we already noted. As $r_* \to \pm \infty$ the effective potential (\ref{eq: potential EM GRAV}) behaves as
\begin{align}  \label{eq: at infinity}
 \lim_{r_* \to + \infty} V(r_*) & = \alpha_E^2, \qquad \qquad \lim_{r_* \to - \infty} V(r_*)  = 0 .
\end{align}
In Fig.\ 1 we plotted the effective potential (\ref{eq: potential EM GRAV}) with $\alpha_E=1$. We get similar plots for other values of the quantity $\alpha_E$. We note in Fig.\ 1 that the effective potentials (\ref{eq: potential EM GRAV}) are step-like.  As in Ref.\ \cite{Myung:2008pr}, for the step-like potential (\ref{eq: potential EM GRAV}) the QNF given in Eq.\ (\ref{eq: QNF GRAV EM}) are purely imaginary.

In Ref.\ \cite{Dotti:2004sh} was proved the linear stability (instability for $D=6$) under tensor perturbations of some higher dimensional spherically symmetric black holes of the Einstein-Gauss-Bonnet theory. They calculated the effective potentials for this type of gravitational perturbations and exploiting the results of Ref.\ \cite{Buell-Shadwick-1995} they showed that these potentials do not support at least one bound state.\footnote{For $D=6$ they showed that the corresponding effective potential support at least one bound state. See also Refs.\ \cite{Gleiser:2005ra}, \cite{Gibbons:2002pq}.}

For the five-dimensional black hole (\ref{eq: metric dilatonic}) we cannot use a similar method to that of Ref.\ \cite{Dotti:2004sh} to show the existence of bound states because the effective potential (\ref{eq: potential EM GRAV}) has a limit different from zero as $r_* \to + \infty $. Thus we cannot use the results of Buell and Shadwick given in Ref.\ \cite{Buell-Shadwick-1995}, because they assumed that the potential goes to zero as the independent variable goes to infinity. Furthermore, to prove the existence of at least one bound state for a given potential we need to calculate the following integral (for more details see Ref.\ \cite{Buell-Shadwick-1995})
\begin{equation} \label{eq: integral}
 \int_{-\infty}^{+\infty} V(r_*) \,\dd r_*, 
\end{equation}  
and depending on its value we know if the effective potential $V(r_*)$ has at least one bound state. For the effective potential (\ref{eq: potential EM GRAV}) the integral (\ref{eq: integral}) is divergent.

\begin{figure}[th]
\begin{center}
\label{figure1}
\psfrag{r}{$r_*$}
\psfrag{V}{$V(r_*)$}
\includegraphics[scale=.4,clip=true]{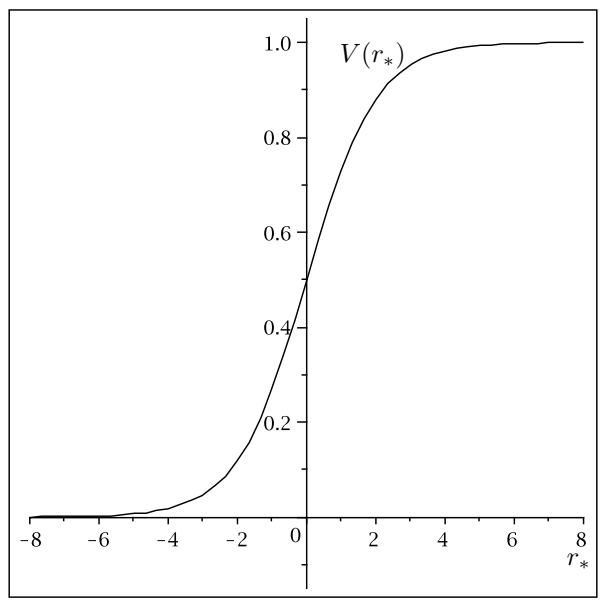}
\caption{ Plot of the effective potential $V$ given in Eq.\ (\ref{eq: potential EM GRAV}) where we take $\alpha_E=1$.} 
\end{center}
\end{figure}

For the non-minimally coupled to gravity massive Klein-Gordon field we find that effective potential in the Sch\"odinger type equation is equal to
\begin{equation} \label{eq: potential Klein}
 V_{KG} (r_*) =  \frac{ \alpha_{KG}^2 \textrm{e}^{r_*}}{\textrm{e}^{r_*} + 1} +  \frac{\xi \textrm{e}^{r_*}}{(\textrm{e}^{r_*} + 1)^2} ,
\end{equation} 
where
\begin{equation}
 \alpha_{KG}^2 = \frac{l(l+2) + 6 \xi }{r_0^2} + \mu^2 .
\end{equation} 
As $r_* \to \pm \infty$ this potential behaves as the potential $V$ of Eq.\ (\ref{eq: potential EM GRAV}) (see formulas (\ref{eq: at infinity})); thus the plot of the potential $V_{KG}$ of (\ref{eq: potential Klein}) is similar to that draw in Fig.\ 1. In a similar way to the potential $V$ of formula (\ref{eq: potential EM GRAV}), for $ V_{KG}$ the integral (\ref{eq: integral}) is divergent.

\section{QNMs of the massive Dirac field}
\label{sect 3}

As is well known, in a $D$-dimensional $G_{D-2}$-symmetric spacetime whose line element we write in the form \cite{Kodama:2003jz}
\begin{equation} \label{eq: general metric}
\dd s^2 =  F(x)^2 \dd t^2 - G(x)^2 \dd x^2 - H(x)^2 \dd \Sigma^2_{D-2},
\end{equation} 
the massive Dirac equation $ i \dirac \psi = \mu \psi$ simplifies to a pair of coupled partial differential equations in two variables \cite{Cho:2007zi,Cho:2007ce,Gibbons:1993hg}
\begin{align} \label{eq: Dirac equation general}
 \partial_t  \psi_2 - \frac{F}{G} \partial_x  \psi_2 & =  \left( i \kappa \frac{F}{H} -  i \mu F \right) \psi_1, \nonumber \\
 \partial_t \psi_1 + \frac{F}{G} \partial_x \psi_1 & =  - \left( i \kappa \frac{F}{H} +  i \mu F \right) \psi_2,  
\end{align} 
where $\kappa$ denotes the eigenvalues of the Dirac operator on the manifold with line element $\dd \Sigma^2_{D-2}$, the functions $\psi_1$ and $\psi_2$ are the components of a two-dimensional spinor $\psi_{2D}$ in the $(t,x)$ sector of the line element (\ref{eq: general metric})
\begin{equation} \label{eq: spinor two-dimensional}
 \psi_{2D}=\left( \begin{array}{c} \psi_1 \\ \psi_2 \end{array} \right).
\end{equation}  
For more details see Refs.\ \cite{Cho:2007zi}, \cite{Cho:2007ce}, \cite{Gibbons:1993hg}. For a different method see Refs.\ \cite{Cotaescu:1998ay}. 

In the present section, by using Eqs.\ (\ref{eq: Dirac equation general}) we compute the QNF of the massive Dirac field propagating in the five-dimensional black hole (\ref{eq: metric dilatonic}). The QNF of the Dirac field evolving in higher dimensional spacetimes have not been studied extensively as for other fields, we know only the results reported in Refs.\ \cite{LopezOrtega:2007sr}, \cite{Cho:2007zi}, \cite{Cho:2007ce}, \cite{Chakrabarti:2008xz}.

As the line element of the five-dimensional black hole (\ref{eq: metric dilatonic}) is of the same type as the line element of formula (\ref{eq: general metric}) we identify the functions $F$, $G$, and $H$ as follows
\begin{equation}
 F = \frac{1}{G} = (1 - \textrm{e}^{-x})^{1/2}, \qquad H=r_0,
\end{equation} 
and therefore in the five-dimensional black hole (\ref{eq: metric dilatonic}) Eqs.\ (\ref{eq: Dirac equation general}) are written as
\begin{align} \label{eq: Dirac equation dilatonic}
 \partial_t \psi_2 - (1 - \textrm{e}^{-x}) \partial_x \psi_2  & =   (1 - \textrm{e}^{-x})^{1/2} \left(\frac{i \kappa}{r_0} - i \mu \right) \psi_1, \nonumber  \\
 \partial_t \psi_1 + (1 - \textrm{e}^{-x}) \partial_x \psi_1 & = - (1 - \textrm{e}^{-x})^{1/2}  \left( \frac{i \kappa}{r_0} + i \mu \right) \psi_2 , 
\end{align} 
where $\kappa$ are the eigenvalues of the Dirac operator on the unit three-dimensional sphere, that is, $\kappa = \pm i(l + \tfrac{3}{2})$, with $l=0,1,2,\dots$, \cite{Camporesi:1995fb}. In the following we only consider $\kappa =  i(l + \tfrac{3}{2})$ (we expect to get similar results for $\kappa = - i(l + \tfrac{3}{2})$).

Now we take the components $\psi_1$ and $\psi_2$ of the two-dimensional spinor (\ref{eq: spinor two-dimensional}) as
\begin{align} \label{eq: psi 1 2 ansatz}
 \psi_1(x,t) = R_1(x)\, \textrm{e}^{-i \omega t } , \nonumber \\
\psi_2(x,t) = R_2(x)\, \textrm{e}^{-i \omega t },
\end{align} 
also, as in Chandrasekhar's book \cite{Chandrasekhar book}, we define the quantity $\theta$ by 
\begin{equation} \label{eq: theta Nariai}
 \theta = \arctan \left( \frac{\mu}{\lambda}\right),
\end{equation} 
where
\begin{equation} \label{eq: lambda dilatonic}
 \lambda = - \frac{i \kappa}{r_0} = \frac{l + \tfrac{3}{2}}{r_0},
\end{equation} 
and taking $R_1 = \textrm{e}^{-i\theta/2} \tilde{R}_1$ and $R_2 = \textrm{e}^{i\theta/2} \tilde{R}_2$ we find that Eqs.\ (\ref{eq: Dirac equation dilatonic}) become
\begin{align} \label{eq: radial three dilatonic}
 (1 - \textrm{e}^{-x})^{1/2}\frac{\dd \tilde{R}_2}{\dd x} + \frac{i \omega \tilde{R}_2}{(1 - \textrm{e}^{-x})^{1/2}}   = \alpha_D \tilde{R}_1,  \nonumber \\ 
 (1 - \textrm{e}^{-x})^{1/2}\frac{\dd \tilde{R}_1}{\dd x} - \frac{i \omega \tilde{R}_1}{(1 - \textrm{e}^{-x})^{1/2}}  = \alpha_D \tilde{R}_2,
\end{align}
where $\alpha_D^2 = (l + \tfrac{3}{2})^2/r_0^2 + \mu^2$. Note that the quantity $\alpha_D^2$ is always different from zero (even when $\mu = l = 0$).

From Eqs.\ (\ref{eq: radial three dilatonic}) we obtain the decoupled ordinary differential equations for the functions $\tilde{R}_1$ and $\tilde{R}_2$
\begin{align} \label{eq: radial dilatonic}
 (1 - \textrm{e}^{-x}) \frac{\dd^2 \tilde{R}_1}{\dd x^2} + \frac{\textrm{e}^{-x}}{2} \frac{\dd \tilde{R}_1}{\dd x} + \frac{i \omega \textrm{e}^{-x} \tilde{R}_1}{2 (1 - \textrm{e}^{-x})} + \frac{\omega^2 \tilde{R}_1}{1 - \textrm{e}^{-x}} - \alpha_D^2 \tilde{R}_1 = 0, \nonumber \\ 
 (1 - \textrm{e}^{-x}) \frac{\dd^2 \tilde{R}_2}{\dd x^2} + \frac{\textrm{e}^{-x}}{2} \frac{\dd \tilde{R}_2}{\dd x} - \frac{i \omega \textrm{e}^{-x} \tilde{R}_2}{2 (1 - \textrm{e}^{-x})} + \frac{\omega^2 \tilde{R}_2}{1 - \textrm{e}^{-x}} - \alpha_D^2  \tilde{R}_2 = 0.
\end{align}
Making the change of variable given in formula (\ref{eq: z coordinate dilatonic}) we find that Eqs.\ (\ref{eq: radial dilatonic}) transform into
\begin{align} \label{eq: dilatonic radial R1 R2}
 \frac{\dd^2 \tilde{R}_1 }{\dd z^2}  & + \left( \frac{1}{2z}- \frac{1}{1-z} \right) \frac{\dd \tilde{R}_1}{\dd z}   +  \frac{\omega^2 \tilde{R}_1}{z^2 (1-z)^2} + \frac{i \omega \tilde{R}_1}{2 z^2 (1-z)} - \frac{\alpha_D^2 \tilde{R}_1}{z(1-z)^2}  = 0 , \nonumber \\
\frac{\dd^2 \tilde{R}_2}{\dd z^2}  & + \left(\frac{1}{2z}- \frac{1}{1-z} \right) \frac{\dd \tilde{R}_2}{\dd z}   + \frac{\omega^2 \tilde{R}_2}{z^2 (1-z)^2} - \frac{i \omega \tilde{R}_2}{2 z^2 (1-z)} - \frac{\alpha_D^2 \tilde{R}_2}{z(1-z)^2}  = 0 .
\end{align}  
These equations are similar to Eq.\ (\ref{eq: dilatonic EM GRAV}) and using the same procedure of Sect.\ \ref{sect 2} we solve the previous two equations. We first solve the equation for the function $\tilde{R}_1$ to calculate the QNF of the component $\psi_1$. Next we present the corresponding results for the component $\psi_2$.

To solve the ordinary differential equation for the function $\tilde{R}_1$ (\ref{eq: dilatonic radial R1 R2}) we make the ansatz of Eq.\ (\ref{eq: Em GRAV ansatz}) where (replace $C$ by $C_1$, $B$ by $B_1$, and $Q$ by $Q_1$)  
\begin{align} \label{eq. B C R1 dilaton black hole}
C_1 = & \left\{ \begin{array}{l} i \omega   \\ \\ - i \omega + \frac{1}{2} \end{array}\right. ,\qquad \quad
B_1 =  \left\{ \begin{array}{l} i \sqrt{\omega^2 - \alpha_D^2}  \\ \\ - i \sqrt{\omega^2 - \alpha_D^2}   \end{array}\right. ,
\end{align}
to find that the function $Q_1$ is a solution to the hypergeometric differential equation (\ref{e: hypergeometric differential equation}) with parameters $a_1$, $b_1$, and $c_1$ equal to
\begin{align} \label{eq: constants hypergeometric dilatonic}
a_1 & = B_1 + C_1 + \tfrac{1}{2}, \nonumber \\
b_1 & = B_1 + C_1 ,  \nonumber \\
c_1 & = 2 C_1 + \tfrac{1}{2} .
\end{align}

To compute the QNF of the massive Dirac field we choose the quantities $C_1$ and $B_1$ as $C_1=i\omega$ and $B_1=i \sqrt{\omega^2 - \alpha_D^2}$. Hence the function $R_1$ is equal to
\begin{align}
 R_1 & =  \textrm{e}^{-i \theta / 2} z^{ i \omega }  (1-z)^{B_1} \left[ \mathbb{D}_1 \,  {}_{2}F_{1}(a_1,b_1;c_1;z) \right.  \nonumber \\
&\left.  + \mathbb{E}_1 \, z^{1/2- 2 i \omega} {}_{2}F_{1}(a_1-c_1+1,b_1-c_1+1;2-c_1;z) \right],
\end{align} 
where $\mathbb{D}_1$ and $ \mathbb{E}_1$ are constants. From the formulas (\ref{eq: dilatonic relations coordinates}), in order to have a purely ingoing wave near the event horizon we must take $\mathbb{D}_1 = 0$. Therefore the function $R_1$ becomes
\begin{align} \label{eq: radial dilatonic R1}
 R_1 & = \mathbb{E}_1 \textrm{e}^{-i \theta / 2} z^{1/2 - i\omega} (1-z)^{B_1} \, {}_{2}F_{1}(a_1-c_1+1,b_1-c_1+1;2-c_1;z) \nonumber \\
& = \mathbb{E}_1 \textrm{e}^{-i \theta / 2}  z^{1/2 - i\omega} (1-z)^{B_1} \, {}_{2}F_{1}(\alpha_1,\beta_1;\gamma_1;z).
\end{align}

If the quantity $\gamma_1 - \beta_1 - \alpha_1 $ is not integer, using the formulas (\ref{eq: dilatonic relations coordinates}) and (\ref{e: hypergeometric property z 1-z}) we get that near infinity the function $R_1$ has a behavior similar to that given in formula (\ref{eq: EM GRAV near infinity}) for the radial function of the boson fields studied in Sect.\ \ref{sect 2}. From the results of Ref.\ \cite{Ohashi:2004wr} and Eqs.\ (\ref{eq: psi 1 2 ansatz}), to satisfy the boundary condition of the QNMs near infinity we must impose the condition
\begin{equation}
 \gamma_1 -\alpha_1 = -n_1, \qquad \textrm{or} \qquad \gamma_1 - \beta_1 = -n_1, 
\end{equation} 
where $n_1=0,1,2,\dots$. From these equations we find that the QNF for the component $\psi_1$ of the massive Dirac field are equal to
\begin{align} \label{eq: QN frequencies dilatonic}
 \omega = - \frac{i}{2(n_1 + \tfrac{1}{2})}\left((n_1 + \tfrac{1}{2})^2 - \alpha_D^2\right), \qquad
 \omega = - \frac{i}{2(n_1 + 1 )}\left((n_1 + 1)^2 - \alpha_D^2\right).
\end{align} 

Notice that for the five-dimensional black hole (\ref{eq: metric dilatonic}) the imaginary part of the QNF for the Dirac field depend on its mass. As in the previous section, the time dependence of the massive Dirac fields is of the form $\exp(- i \omega t)$, hence we need $\im (\omega) < 0$ in order that the perturbations decay in time. Thus the QNMs with frequencies (\ref{eq: QN frequencies dilatonic}) have amplitudes increasing with time when the quantities $\alpha_D$ and $n_1$ satisfy
\begin{equation}
\alpha_D > n_1 + \tfrac{1}{2}.
\end{equation} 
 
In a similar way to the component $\psi_1$, by imposing the boundary conditions of the QNMs we find that the QNF for the component $\psi_2$ are
\begin{align} \label{eq: QN frequencies dilatonic R2}
 \omega  = - \frac{i}{2(n_2 + \tfrac{1}{2})}\left((n_2+\tfrac{1}{2})^2 - \alpha_D^2\right), \qquad
 \omega  = - \frac{i}{2n_2}\left(n_2^2 - \alpha_D^2\right),
\end{align} 
where $n_2 = 0,1,2,\dots$ for the first set of QNF and $n_2 = 1,2,\dots$, for the second set. These frequencies determine QNMs with amplitudes decreasing in time when the mode number $n_2$ and the quantity $\alpha_D$ fulfill the condition $n_2 > \alpha_D$.

Thus depending on the values of $\mu$, $l$, $r_0$, and $n_1$ (or $n_2$) our results for the QNF of the massive Dirac field given in formulas (\ref{eq: QN frequencies dilatonic}) and (\ref{eq: QN frequencies dilatonic R2}) show the existence of some unstable QNMs (those with $\alpha_D > n_1 + 1/2$ for the component $\psi_1$ and those with $\alpha_D > n_2$ for the component $\psi_2$). This result is similar to that of the previous section for the electromagnetic and Klein-Gordon perturbations and different from that of Ref.\ \cite{Becar:2007hu}.

As for the bosons fields of Sect.\ \ref{sect 2}, Eqs.\ (\ref{eq: radial dilatonic}) of the massive Dirac field simplify to Schr\"odinger type equations with effective potentials equal to \cite{Chandrasekhar book}
\begin{equation} \label{eq: potential Dirac}
 V_{\pm}(r_*) = \frac{\alpha_D^2 \textrm{e}^{r_*}}{\textrm{e}^{r_*} + 1} \pm \frac{\alpha_D}{2}  \frac{ \textrm{e}^{r_*/2}}{(\textrm{e}^{r_*} + 1)^{3/2}}.
\end{equation}
Similar to the effective potentials (\ref{eq: potential EM GRAV}) and (\ref{eq: potential Klein}), as $r_* \to \pm \infty$ the effective potentials (\ref{eq: potential Dirac}) behave as
\begin{align}
\lim_{r_* \to + \infty} V_{\pm}(r_*) & = \alpha_D^2, \qquad \qquad \lim_{r_* \to - \infty} V_{\pm}(r_*)  = 0 .
\end{align}  
In Fig.\ 2  we plot the potentials $V_+$ and $V_-$ for $\alpha_D = 1$. Notice that the effective potential $V_-$ is negative for sufficiently large negative values of $r_*$. As for the effective potentials of Eqs.\ (\ref{eq: potential EM GRAV}) and (\ref{eq: potential Klein}), the integral (\ref{eq: integral}) diverges for the effective potentials $V_{\pm}$ of the massive Dirac field given in formula (\ref{eq: potential Dirac}). Thus using the method of Ref.\ \cite{Dotti:2004sh} we cannot study the linear stability of the black hole (\ref{eq: metric dilatonic}) under massive Dirac fields.

\begin{figure}[th]
\begin{center}
\label{figure2}
\psfrag{r}{$r_*$}
\psfrag{V1}{$V_+$}
\psfrag{V2}{$V_-$}
\includegraphics[scale=.4,clip=true]{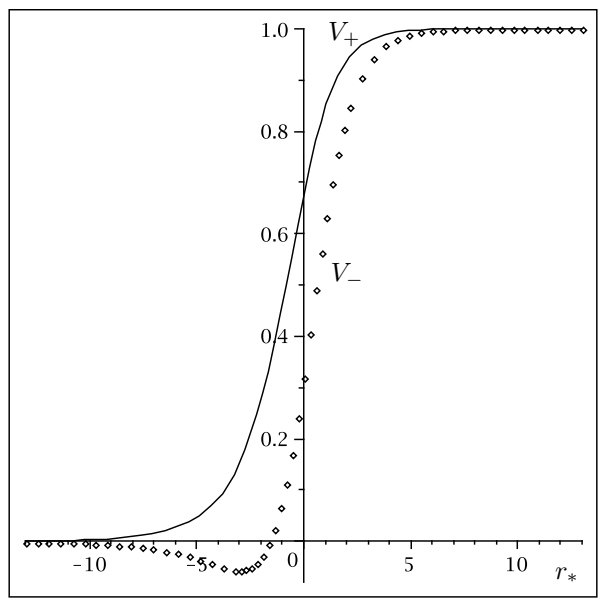}
\caption{ Plots of the effective potentials $V_+$ (solid line) and $V_-$ (broken line) given in Eq.\ (\ref{eq: potential Dirac}), where we take $\alpha_D=1$.} 
\end{center}
\end{figure}

\section{Discussion}
\label{sect 4}

Our results (\ref{eq: QNF GRAV EM}), (\ref{eq: QNF Klein Gordon}), (\ref{eq: QN frequencies dilatonic}), (\ref{eq: QN frequencies dilatonic R2}) for the QNF of the five-dimensional dilatonic black hole (\ref{eq: metric dilatonic}) show that there are unstable modes satisfying the boundary conditions of the QNMs. Therefore due to existence of these unstable QNMs the five-dimensional spacetime (\ref{eq: metric dilatonic}) is linearly unstable. This result is different from that reported in Ref.\ \cite{Becar:2007hu}. 

We remark that in the five-dimensional black hole (\ref{eq: metric dilatonic}) the QNF of the different fields (massive or massless) have the same mathematical form (see our formulas (\ref{eq: QNF GRAV EM}), (\ref{eq: QNF Klein Gordon}), (\ref{eq: QN frequencies dilatonic}), (\ref{eq: QN frequencies dilatonic R2})).

We believe that using other methods we must settle the problem of the classical stability or instability of the five-dimensional studied in this paper. In particular, to investigate if some extension of the method exploited in Ref.\ \cite{Dotti:2004sh} can be applied to determine if this black hole is classically stable or unstable deserve some attention. If we confirm the classical instability of this black hole, the implications of this fact on the calculations made in this spacetime deserve further study. 

Notice that the analysis of Sects.\ \ref{sect 2} and \ref{sect 3} does not include gravitational perturbations because these are coupled with those of the dilaton field. Thus we expect that the equations of motion for the three types of gravitational perturbations cannot be reduced to partial differential equations similar to Eq.\ (\ref{eq: partial differential fields}). Therefore for the gravitational perturbations we believe that the analysis is more difficult than for the fields studied in Sects.\ \ref{sect 2} and \ref{sect 3}. 

As is well known the effective potential for the scalar type gravitational perturbations is more complicated than those for the vector and tensor type gravitational perturbations \cite{Kodama:2003kk,Kodama:2003jz}. This fact implies that the linear stability under scalar type gravitational perturbations  of several higher dimensional asymptotically de Sitter and anti-de Sitter static black holes is not proved in the present \cite{Ishibashi:2003ap}. Also from the recent results of Refs.\ \cite{Konoplya:2007jv} the computation of the QNF for the scalar type gravitational perturbations propagating in the five-dimensional black hole (\ref{eq: metric dilatonic}) is a relevant issue, since some higher dimensional black holes are linearly unstable under this type of gravitational perturbations. Thus we believe that the linear stability of the black hole (\ref{eq: metric dilatonic}) under gravitational perturbations is worth to be considered in order to prove or disprove our results.

For the five-dimensional black hole (\ref{eq: metric dilatonic}), from the results of the present work and of Ref.\ \cite{Becar:2007hu} for the vector and scalar type electromagnetic fields, the massive Dirac field and the massive Klein-Gordon field for $\xi < \tfrac{1}{4}$, we get that the real part of the QNF is zero. This result indicates that Hod's conjecture \cite{Hod:1998vk} is not valid for the black hole (\ref{eq: metric dilatonic}), because in this conjecture we need to assume that the real part of the QNF is different from zero and depends only on the physical parameters of the black hole.

\appendix
\section{Quasinormal modes of a two-dimensional black hole}
\label{app 1}

In Ref.\ \cite{Becar:2007hu} were computed the QNF of the non-minimally coupled to gravity Klein-Gordon field propagating in the uncharged two-dimensional black hole of Ref.\ \cite{McGuigan:1991qp} with line element (the $(t,x)$ sector of the metric (\ref{eq: metric dilatonic}))
\begin{equation} \label{eq: metric dilatonic two}
{\rm d} s^2 = (1 - \textrm{e}^{-x})\, {\rm d} t^2 - \frac{ {\rm d} x^2 }{ (1 -  \textrm{e}^{-x}) } .
\end{equation}  
The result obtained in Ref.\ \cite{Becar:2007hu} for the QNF of the massive Klein-Gordon field with non-minimal coupling to gravity is (Eq.\ (25) in Ref.\ \cite{Becar:2007hu}) 
\begin{equation} \label{eq: QNF Becar 2D}
 \omega = -\frac{i}{4} \left[ 1 - \sqrt{1-4\zeta}  - \frac{(1 +  \sqrt{1-4\zeta} ) \mu^2}{n^2 + n + \zeta} + n\left(2 -\frac{2 \mu^2}{n^2 + n +\zeta} \right)\right] ,
\end{equation} 
where $n$ is the mode number ($n=0,1,2,\dots$), $\mu$ is the mass of the Klein-Gordon field, and $\zeta$ is the coupling constant between the Klein-Gordon field and the scalar curvature of the spacetime (\ref{eq: metric dilatonic two}).

Here we note only the following facts. It is possible to write the QNF (\ref{eq: QNF Becar 2D}) in the simpler form
\begin{equation} \label{eq: QNF 2D KG 1}
 \omega = -\frac{i}{4} \left[ 2n + 1 -   \sqrt{1-4\zeta} - \frac{4 \mu^2}{2n + 1 - \sqrt{1 - 4 \zeta}}\right] ,
\end{equation} 
whose mathematical form is similar to that of the QNF (\ref{eq: QNF Klein Gordon}) of the five-dimensional black hole (\ref{eq: metric dilatonic}). Also Becar, et al.\ do not report the following QNF 
\begin{equation} \label{eq: QNF 2D KG 2}
 \omega = -\frac{i}{4} \left[ 2n + 1 +  \sqrt{1-4\zeta} - \frac{4 \mu^2}{2n + 1 + \sqrt{1 - 4 \zeta}}\right] .
\end{equation} 

In Ref.\ \cite{Becar:2007hu} were imposed two conditions to find the QNF of the Klein-Gordon field (Eqs.\ (23) and (24) of the previous reference, $a=-n$ and $b=-n$, respectively), but they report only the QNF (\ref{eq: QNF Becar 2D}). Owing to the parameters $a$ and $b$ are different (see Eqs.\ (11) and (12) of Ref.\ \cite{Becar:2007hu}), we expect to get two different sets of QNF (Eqs.\ (\ref{eq: QNF 2D KG 1}) and (\ref{eq: QNF 2D KG 2}) above).

Moreover Eqs.\ (\ref{eq: QNF 2D KG 1}) and (\ref{eq: QNF 2D KG 2}) allow us to discuss the stability of the two-dimensional black hole (\ref{eq: metric dilatonic two}) in a simpler way. Thus for $\zeta \leq \tfrac{1}{4}$, the two-dimensional black hole  (\ref{eq: metric dilatonic two}) is linearly unstable if $n$, $\zeta$, and $\mu$ satisfy
\begin{equation} \label{eq: instability condition 2D}  
 n < \mu \pm \frac{\sqrt{1 - 4 \zeta}}{2} - \frac{1}{2}.
\end{equation}  
 
Using Eqs.\ (\ref{eq: QNF 2D KG 1}), (\ref{eq: QNF 2D KG 2}), and (\ref{eq: instability condition 2D}) we reach the same conclusions on the linear instability of the two-dimensional black hole (\ref{eq: metric dilatonic two}) that those already published in Ref.\ \cite{Becar:2007hu} (see also Refs.\ \cite{AzregAinou:1999kb}, \cite{Frolov:2000jh}, \cite{Zelnikov:2008rg}). For $\zeta > \tfrac{1}{4}$ also we find similar results to those reported in Ref.\ \cite{Becar:2007hu}.

\section{Acknowledgments}

I thank Dr.\ C.\ E.\ Mora Ley, Dr.\ R.\ Garc\'{\i}a Salcedo, and  Dr.\ O.\ Pedraza Ortega for their interest in this paper. This work was supported by CONACYT M\'exico, SNI M\'exico, EDI-IPN, COFAA-IPN, and Research Project SIP-20080794.

\end{document}